# The development of the architecture of Distributed Network Intrusion Detection System (D-NIDS)

A.Bushev, S.Vlasenko, I.Glotov, Y.Monakhov, A.Tishin

Vladimir State University named after Alexander and Nikolay Stoletovs

Nowadays corporate networks of data transmission are an important constituent of an organization's infrastructure. A consolidation of computers in a network allows to provide resource sharing of the network and an operational access to all corporate information, organize a high-speed access of users to Internet and create reliable centralized resources of backup and holding the information. Robert Metcalf, participated in the creation of Ethernet, was nearly the first who gave the quantitative estimation of a networks value. According to his estimate, "value" of a network, in all senses, is proportional to the square of a number of nodes in it. [1]

D-NIDS is one of the obligatory components of an efficient architecture of provision the information security in a multilevel security arrangement of a corporate network of data transmission. [3] As IDS (Intrusion-Detection System) is the main way of monitoring the condition of a network and analysis of its immunity, as well as provides a protection of data systems and individual nodes of a network from internal and external attacks.

It is usually picked out three main subsystems in IDS: sensor subsystem, subsystem of analysis and interface. [2]

Network IDSs get an access to network traffic, attaching to a hub or router which adapted for mirroring of ports (mirror/span-port) or network splitters (TAP – devices). Both these ways have their own disadvantages: when using mirroring – it is a loss of packets in the mirror/span-port, but when using network splitters it is necessary to have a large amount of them for traffic analysis of the whole network.

In the developed system it is supposed to realize the sensor subsystem is based on using a new way of rerouting traffic to sensors of net IDS, using the features of link level addressing of ISO/OSI. [3] This way is based on the modification of APR-tables of endpoint devices, that allows one sensor to control traffic from several nodes inside the broadcast segment of network, also it is devoid of disadvantages of two previous ways.

The scheme of the integration of the developed IDS into a corporate network is represented on the figure 1.

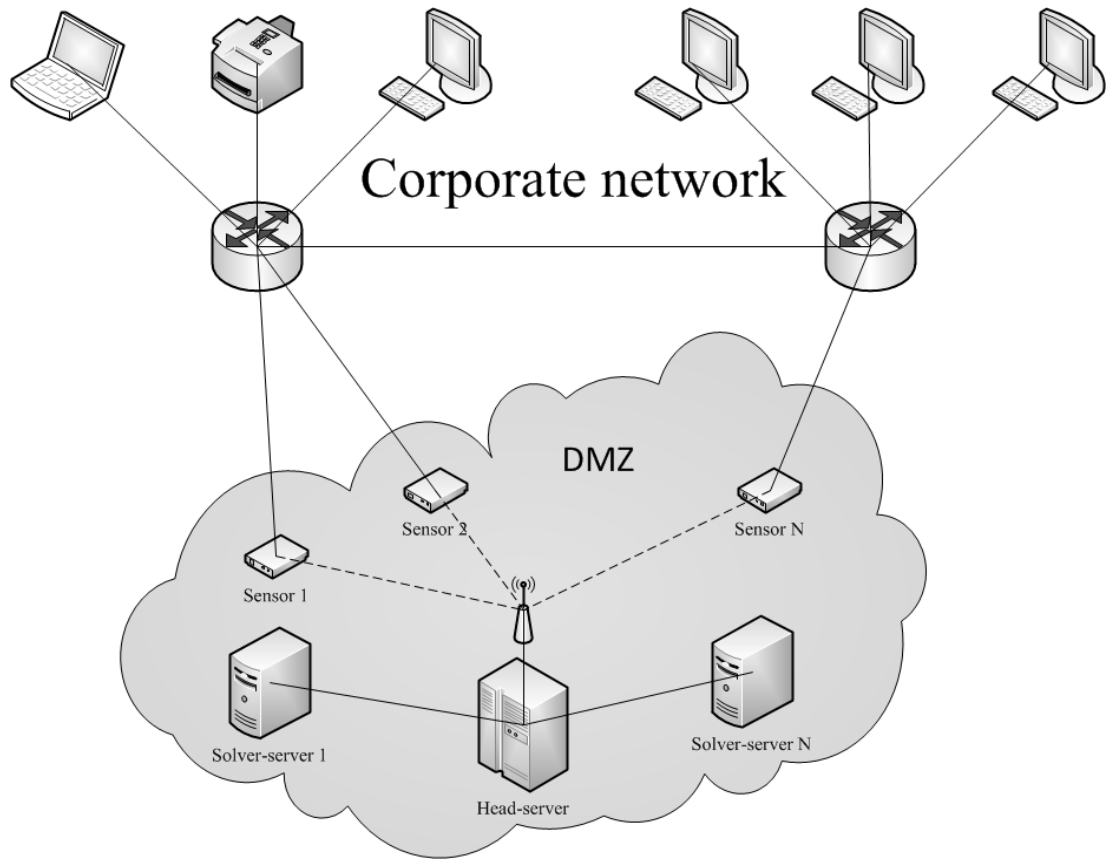

Figure 1

In the system we plan to use computer appliance: sensors, a head-server, solver-servers.

It is planned to use in the system the following hard-software devices: sensors, a head-server, solver-servers.

Sensors are represented by devices, attached to switching equipment and used for gathering traffic. A hardware component is based on a base card of BeagleBone, that possesses the following necessary attributes: the ARM architecture of the processor, an availability of USB host and Ethernet-port. Under this architecture an Linux-based OS was developed, which provides the dedicated software. This software performs the capture, balancing and transmission of traffic. Sensors interact with each other and the head-server inside the demilitarized zone, formed by dint of a protected wi-fi connection.

Hardware attributes of the head-server are depend on the amount of network and volume of traffic transferred by it. As a software linux-base OS and a dedicated software are used, which is used for receiving, processing, storage and transferring traffic. The storage of traffic is realized in the document-centric database MongoDB, and messages from solver-servers are stored in the relational database PostgreSQL.

Solver-servers encapsulate into themselves modules of traffic analysis in order to detect attacks and suspicious actions. Thus hardware attributes of a specific solver-server depend on implemented analysis algorithms and traffic volume. Architectural features of a solver-server allow to put it not only on an individual computer but on a head-server as well. A format of messages modules traffic analysis must conform to IDMEF (The Intrusion Detection Message Exchange Format), described in RFC4765. A standardization of message format will allow to use traffic analysis modules of another developers and make writing of a new modules easier.